\documentclass[]{appolb}
\usepackage{graphicx}
\usepackage{amsmath}
\usepackage[hyperref]{xcolor}
\usepackage[acronym]{glossaries-extra}
\usepackage{cleveref}
\setabbreviationstyle[acronym]{long-short}
\newacronym{ff}{FF}{four-fermion}
\newacronym[plural=QFTs, longplural=quantum field theories]{qft}{QFT}{quantum field theory}
\newacronym{qcd}{QCD}{Quantum Chromodynamics}
\newacronym{qsl}{QSL}{quantum spin liquid}
\newacronym{cep}{CEP}{critical end point}
\newacronym{lp}{LP}{Lifshitz point}
\newacronym{njl}{NJL}{Nambu-Jona-Lasinio}
\newacronym{qm}{QM}{quark-meson}
\newacronym{gn}{GN}{Gross-Neveu}
\newacronym{chign}{$\chi$HGN}{chiral Heisenberg-Gross-Neveu}
\newacronym{ip}{IP}{inhomogeneous phase}
\newacronym{hbp}{HBP}{homogeneous broken phase}
\newacronym{sp}{OSP}{ordinary symmetry-restored phase}
\newacronym{uv}{UV}{ultra-violet}
\glssetcategoryattribute{acronym}{nohyperfirst}{true}

\newcounter{numrefs}
\makeatletter
\newcommand{\Rcite}[1]{%
	\setcounter{numrefs}{0}
	\@for\@temp:=#1\do{\stepcounter{numrefs}}
	\ifnum\value{numrefs}>1%
	Refs.~\cite{#1}%
	\else%
	Ref.~\cite{#1}%
	\fi%
}
\makeatother

\begin{document}
\title{Lattice study of disordering of inhomogeneous condensates and the Quantum Pion Liquid in effective $O(N)$ model%
\thanks{Presented at the \textit{Exited QCD 2024}, Benasque, Spain, 15-19 January, 2024}%
}
\author{Marc Winstel
\address{Institut für Theoretische Physik, Goethe-Universität,
	\\
	Max-von-Laue-Straße 1, D-60438 Frankfurt am Main, Germany.}
\\[3mm]
{Semeon Valgushev 
\address{Department of Physics and Astronomy, Iowa State University,\\ Ames, IA, 50011, USA.}
}
}
\maketitle
\begin{abstract}
In this talk, we study a scalar $O(N)$ model with a so-called moat regime -- a regime with negative bosonic wave function renormalization -- using lattice field theory.
For negative bare wave function renormalization, inhomogeneous condensates are solutions of the classical equations of motions.   	
Using hybrid Monte Carlo simulations we demonstrate how bosonic quantum fluctuations disorder the inhomogeneous condensate.
Instead, one finds a so-called Quantum Pion Liquid, where bosonic correlation functions are spatially oscillating, but also exponentially decaying. 
\end{abstract}
  
\section{Introduction}
Scalar $O(N)$ field theories are a useful tool for the study of phase transitions.
This is especially true in the case of \gls{qcd} at finite density where 1) Lorentz symmetry is broken explicitly and therefore new phases can arise and 2) Monte-Carlo simulations are not available due to the sign problem. In particular, QCD might exhibit the so-called \gls{ip} -- where in addition to chiral symmetry also translational invariance is spontaneously broken  \cite{Buballa:2014tba}. In order to study this phenomenon one can consider effective models such as \gls{njl} \cite{Koenigstein:2021llr} and scalar $O(N)$ models \cite{Pisarski:2020dnx}. Instead (or in addition to) of critical end-point of $O(4)$ universality class one might expect to find so-called Lisfhitz point where three phases meet: disordered, ordered and the IP.

In previous studies of an \gls{njl}-type model in $1+1$ dimensions it was found that the IP coincides with so-called moat regime \cite{Koenigstein:2021llr}, where the dispersion relation of quasi-particles exhibits minimum at non-zero momentum. This unusual feature manifests itself as periodic spatial oscillations of two-point correlation functions. However, the connection of the moat regime to IP is not straightforward. Quantum fluctuations are expected to weaken ordered phases such as the \gls{ip} \cite{Lenz:2020bxk, Stoll:2021ori},
and in fact they can even eliminate them. In particular, it was argued that in $3+1$ dimensional scalar $O(N)$ model one can find so-called Quantum Pion Liquid (Q$\pi$L) analogous to Quantum Spin Liquid (QSL) instead of IP \cite{Pisarski:2020dnx} due to transverse fluctuations. In this case, Q$\pi$L is characterized by small yet always finite mass and is accompanied by the moat regime. The Q$\pi$L regime can also be generated through mixing effects of scalar and vector modes \cite{Haensch:2023sig, Winstel:2024dqu}.

In this work we set ourselves to go beyond limitations of previous works and address the problem of inhomogeneous phase using ab-initio lattice calculations of scalar $O(N)$ model at finite $N$.

\section{Model and algorithm}
We consider an $O(N)$ model in $3+1$ dimensions with spatial higher-derivative terms as defined in \cite{Pisarski:2020dnx} and take into account only the static Matsubara mode $\beta \omega_0 = 0$ where $\beta = 1/T$. In this case, an effective $3$-dimensional Lagrangian of the static mode reads as
\begin{equation}
L_{\mathrm{eff}} = \frac{Z}{2} \left(\partial_{j} \vec{\phi}\right)^2 + \frac{1}{2 M^2} \left(\sum_j \partial_j^2 \vec{\phi} \right)^2 +\frac{m^2}{2} \vec{\phi}^{\,2} + \frac{\lambda N}{4} (\vec{\phi}^{\,2})^2. \label{eq:model}
\end{equation}
This Lagrangian can be analyzed in mean-field approximation and in the large-$N$ limit. In the mean field approach, it features a Lifshitz point at $Z=0$ and $m^2=0$ where three phases meet: disordered, ordered and the IP. The ground state of the IP is a chiral spiral
\begin{equation} 
	    \vec{\phi} = \phi_0\left(\cos(k_0 z), \sin(k_0 z),  \phi_\perp = \vec{0}\right)^T
	\end{equation}
 where $k_0$ and $\phi_0$  are such that they minimize the free energy. In contrast, in the large-$N$ limit the Lifshitz point is gone. The IP is replaced by a Q$\pi$L characterized by dynamically generated mass gap and oscillating two-point function for certain regions in the $\left(m^2, Z\right)$ plane. The Q$\pi$L can exist for both the bare parameter $Z < 0$ but also for positive $Z$. As was argued in \Rcite{Pisarski:2020dnx}, even at finite values of $N$ transverse fluctuations should be strong enough to disorder the condensate and transform the IP into a Q$\pi$L.

In order to write down a Lagrangian suitable for numerical calculations we discretize derivatives:
\begin{equation}
	[\Delta_i]_{x,y} = \frac{1}{12a^2} \left( -\delta_{y,x+2e_i} + 16\delta_{y,x+e_i} - 30 \delta_{y,x} + 16 \delta_{y,x-e_i} - \delta_{y,x-2e_i}\right)
\end{equation}
where $e_i$ is a unit lattice step in the $i$-th direction, and perform standard Hybrid Monte-Carlo (HMC) simulations. We run the calculations for $N=1,2,4$ and lattice sizes $V = 12^3, 16^3, 20^3$ and various values of coupling constants. We generated around $3000-7000$ independent configurations for each point of the phase diagram and used Jackknife algorithm for error estimation.

\section{Results}
	As discussed above, one expects different alternative scenarios to an \gls{ip} from analytical approximations to the partition function of model \eqref{eq:model}.
	Both the IP as well as the Quantum Pion Liquid regime are characterized by a particular behavior of bosonic correlations functions. 
	Also, direct access of $\langle \vec{\phi}(x) \rangle$ or $\langle |\vec{\phi}|(x) \rangle$ is not suitable for detecting \glspl{ip} due to destructive interference \cite{Lenz:2020bxk}.
	Thus, a straightforward choice for an observable characterizing these different regimes are the  spatial correlation functions between the bosonic fields $\vec{\phi}$ 
	\begin{equation}
	C^{ij}( \vec x) = \langle c^{ij}(\vec x) \rangle = \tfrac{1}{V} \sum_{\vec{y}}  \langle \phi^i(\vec y+ \vec x) \phi^j(\vec y) \rangle, \label{eq:corr_i_j}
	\end{equation}
	where the sum over lattice sites $y$ is used to get more statistics. 
	In order to characterize the different regimes, we use fits of $C^{jj}$ for the respective regimes:
 \begin{itemize}
     \item Decaying oscillations $C^{ij}( x)\sim \delta_{ij}e^{-m_r x} \cos(k x)$ for the {Q$\pi$L}, \gls{sp}  (using $k = 0$) and the IP (using $m_r = 0$).
     \item Algebraically decaying oscillations $C^{ij}( x)\sim \delta_{ij}\frac{\cos(k x)}{x^{\alpha}} $ for possible quasi-long range order (a.k.a liquid crystal)
 \end{itemize}
 
	As a criterion to compare the fit qualities we use the coefficient of determination.
	In the following, we will present results for $M^2 = 1.0$, $\lambda N = 1.0$ and $V = 20^3$. 
	However, we note that our findings are stable among different volumes $V = 12^3, 16^3, 20^3$ except for very small, negative $Z$, where the decay rates are getting small and larger volumes are needed.
	
	\begin{figure}[t]
		\centering
		\includegraphics[width=.33\textwidth]{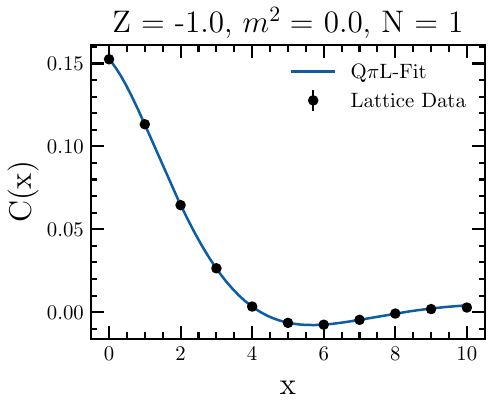} \includegraphics[width=.33\textwidth]{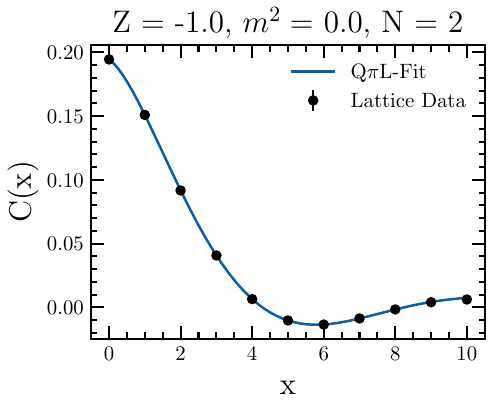}\includegraphics[width=.33\textwidth]{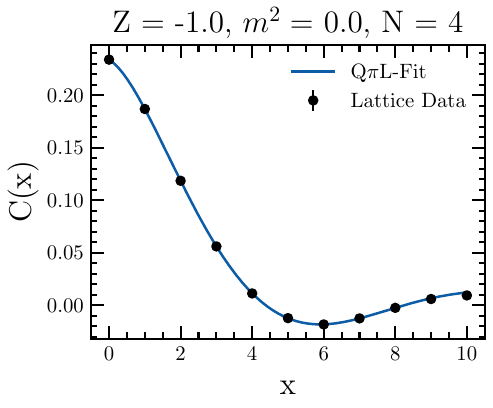}
		\caption{\label{fig:N124_Corr} Plot of $C(x) = C^{11}\left(\left(x, 0,0\right)\right)$ for $Z = -1.0, m^2 = 0.0$. The preferred fit scenario is determined using the coefficient of determination. \textbf{(left)} $N=1$. \textbf{(middle)} $N=2$. \textbf{(right) N = 4}.}
	\end{figure}
	
	In \cref{fig:N124_Corr}, we plot $C^{1 1}$, see \cref{eq:corr_i_j}, as well as the preferred fit scenario for $Z = -1.0, m^2 = 0.0$ and $N = 1, 2, 4$.
	As one can see, the Quantum Pion Liquid fit scenario is preferred independently of $N$.
	From \Rcite{Pisarski:2020dnx}, one would have expected to obtain an \gls{ip} at least for $N= 1$ since there is no disordering through Goldstone modes of $O(N)$ symmery breaking.
	When further decreasing $Z$ the obtained exponential decay rate gets smaller in consistency with the predictions from \Rcite{Pisarski:2020dnx} from the large-$N$ limit.

\begin{figure}[t]
		\begin{center}
		\centering
		\includegraphics[scale=0.9]{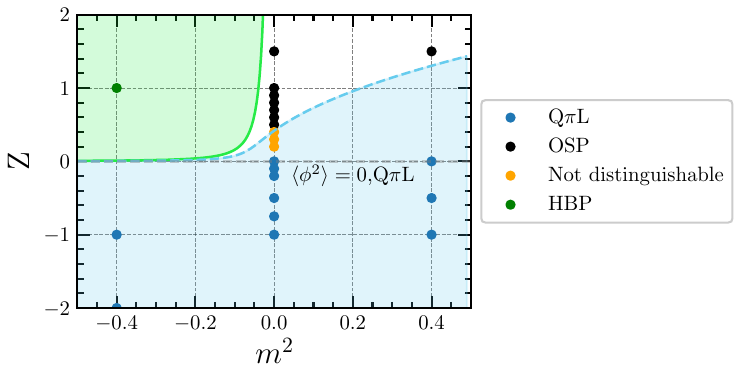}
	\end{center}
	\caption{\label{fig:PhaseDiagN=1} Phase diagram from fits for $N=1$. The green color denotes Homogeneous Broken Phase (HBP), white --- Ordinary Symmetric Phase (OSP) and blue --- Quantum Pion Liquid (Q$\pi$L) regions according to the large-$N$ results from Ref.~\cite{Pisarski:2020dnx}.}
\end{figure}

	In general, the observation of the different regimes in the $\left(m^2, Z\right)$ plane is similar to the large-$N$ findings \cite{Pisarski:2020dnx} for all studied $N$.
	For \cref{fig:PhaseDiagN=1}, we depict the found regimes for $N = 1$ in comparison to the large-$N$ boundary lines. 
	Our findings are in agreement with the large-$N$ prediction, although one has to note that the comparison of different fit scenarios does not yield conclusive results near any of the phase boundaries.
	The phase diagram in the $\left(m^2, Z\right)$ plane seems to be identical for all studied values of $Z, m^2$ and $N$ up to the current status of investigation.

\subsection{Infinite volume limit with the external field}
\begin{figure}[t]
	\includegraphics[width=.48\columnwidth]{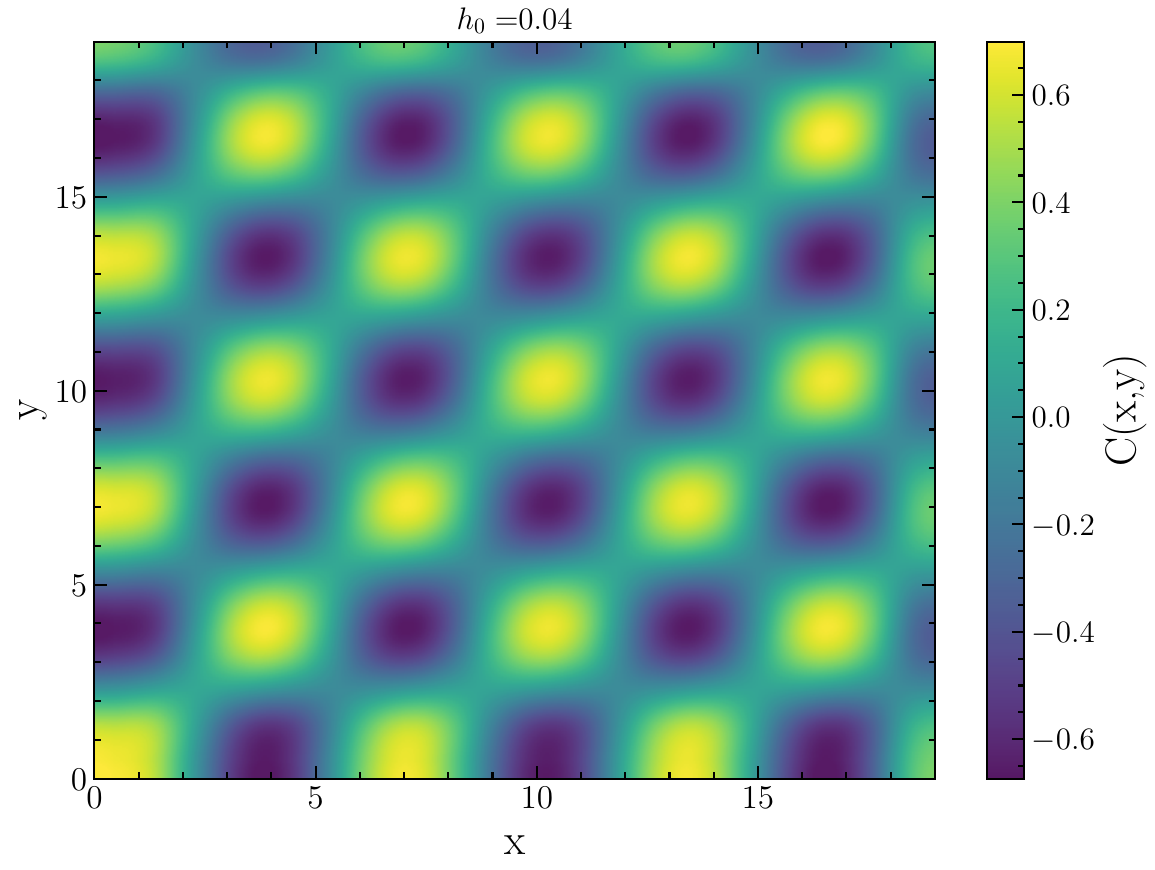}\hfill
	\includegraphics[width=.48\columnwidth]{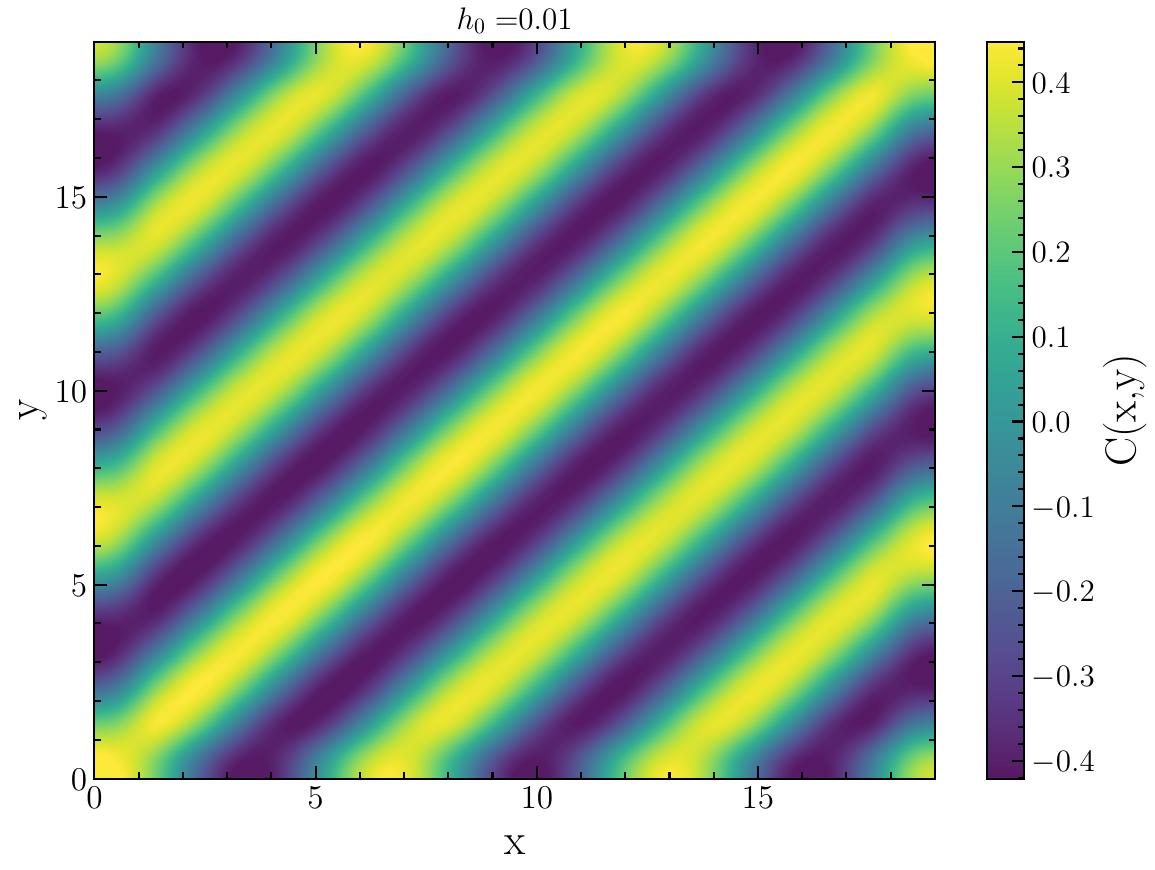}
	\caption{\label{fig:ExtBreak} Color code of $C(0, 0, x, y)$ for $Z = -2.0, m^2 = 0.0$. Note that the scaling of the color bar differs in both plots. \textbf{(left)} $h_0 = 0.04$. \textbf{(right)} $h_0 = 0.01$.}
\end{figure}

The above findings can be seen as an indication that the \gls{ip} might not exist in the $\left(m^2, Z\right)$ phase diagram of \cref{eq:model}.
However, one still has to perform the thermodynamic limit, i.e. $V \rightarrow \infty$, in order to identify phase transitions.
On the lattice we only simulate finite volumes. 
Thus, we are in the need of an extrapolation method of our findings to the infinite volume.
While a finite volume scaling of the exponential decay rate $m_r$ in the Quantum Pion Liquid regime is plagued by fitting and statistical errors, one can go the traditional route of studying phase transition using an external symmetry breaking parameter.
This is introduced for $N=1$ by modifying \cref{eq:model} using an oscillating external field
\begin{equation}
L_{h_0} = L_{\mathrm{eff}} - h(x) \phi_1(x), \quad 
	h(x) = \frac{h_0}{\sqrt{2\pi}L\,\sigma_0 }\sum_{n = 0}^{L-1}  \e^{-\frac{1}{2\sigma} \left(p_n  - k_0\right)^2} \cos(p_n x_2) 
\end{equation}
with Gaussian distributed momenta, where $p_n = 2 \pi n / V^{1/3}$ and $\sigma_0 = 0.1$.
In order to determine the peak of the Gaussian in momentum space, we extract $k_0$ from the simulations with $h_0 = 0$. 
For $Z = -1.0, m^2 = 0.0$ and $V = 20^3$ as used in \cref{fig:ExtBreak} we determine $k_0 =  0.942 \approx 3 \times 2\pi / 20 $.  
Since the introduction of the external symmetry breaking term does not only break translational but also rotational invariance for $h_0 \neq 0$, we expect that the correlation function still depends on the relative differences in the $x_0$ and $x_1$ directions.
Translational symmetry breaking should then have an impact on its dependence on the $x_2$ directions, i.e., one expects
\begin{equation}
    C(x_0 - y_0, x_1 - y_1, x_2, y_2) = C^{11}(x,y).
\end{equation}

In \cref{fig:ExtBreak}, we plot $C(0, 0, x, y)$ as a color map. 
For the left plot with $h_0 = 0.04$, one can directly see the signs of translational symmetry breaking due to external field.
On the other hand, translational invariance is restored for $h_0 = 0.01$ on the right plot of \cref{fig:ExtBreak} since there clearly is only a dependence of $C$ on $x-y$. It is important to determine the transition point $h_{0,c} \in (0.01,0.04)$ between these behaviors as a function of the volume $V$ in order to clarify the fate of translation symmetry.
If an extrapolation of the computed values of $h_{0,c}$ reveals that $h_{0,c}(V) = 0.0$ for $V \rightarrow \infty $, the phase diagram of model \eqref{eq:model} would in turn consist of an IP for $N=1$ in the thermodynamic limit.
Alternatively, one could rigorously establish that inhomogeneous condensates can simply be disordered by the inclusion of bosonic quantum fluctuations.
It would also be interesting to study the dependence of $C^{ij}(x,y)$ on directions which are transverse to $h(x)$ in order to study spontaneous breaking of the rotational symmetry.
This work is planned for a future publication.

\bibliographystyle{JHEP}
\bibliography{literature.bib}

\providecommand{\href}[2]{#2}\begingroup\raggedright\begin{thebibliography}{1}

\bibitem{Buballa:2014tba}
M.~Buballa and S.~Carignano, \emph{{Inhomogeneous chiral condensates}},
  \href{https://doi.org/10.1016/j.ppnp.2014.11.001}{\emph{Prog. Part. Nucl.
  Phys.} {\bfseries 81} (2015) 39}
  [\href{https://arxiv.org/abs/1406.1367}{{\ttfamily 1406.1367}}].

\bibitem{Koenigstein:2021llr}
A.~Koenigstein, L.~Pannullo, S.~Rechenberger, M.J.~Steil and M.~Winstel,
  \emph{{Detecting inhomogeneous chiral condensation from the bosonic two-point
  function in the (1 + 1)-dimensional Gross\textendash{}Neveu model in the
  mean-field approximation*}},
  \href{https://doi.org/10.1088/1751-8121/ac820a}{\emph{J. Phys. A} {\bfseries
  55} (2022) 375402} [\href{https://arxiv.org/abs/2112.07024}{{\ttfamily
  2112.07024}}].

\bibitem{Pisarski:2020dnx}
R.D.~Pisarski, A.M.~Tsvelik and S.~Valgushev, \emph{{How transverse thermal
  fluctuations disorder a condensate of chiral spirals into a quantum spin
  liquid}}, \href{https://doi.org/10.1103/PhysRevD.102.016015}{\emph{Phys. Rev.
  D} {\bfseries 102} (2020) 016015}
  [\href{https://arxiv.org/abs/2005.10259}{{\ttfamily 2005.10259}}].

\bibitem{Lenz:2020bxk}
J.~Lenz, L.~Pannullo, M.~Wagner, B.~Wellegehausen and A.~Wipf,
  \emph{{Inhomogeneous phases in the Gross-Neveu model in 1+1 dimensions at
  finite number of flavors}},
  \href{https://doi.org/10.1103/PhysRevD.101.094512}{\emph{Phys. Rev. D}
  {\bfseries 101} (2020) 094512}
  [\href{https://arxiv.org/abs/2004.00295}{{\ttfamily 2004.00295}}].

\bibitem{Stoll:2021ori}
J.~Stoll, N.~Zorbach, A.~Koenigstein, M.J.~Steil and S.~Rechenberger,
  \emph{{Bosonic fluctuations in the $( 1 + 1 )$-dimensional
  Gross-Neveu(-Yukawa) model at varying $\mu$ and $T$ and finite $N$}},
  \href{https://arxiv.org/abs/2108.10616}{{\ttfamily 2108.10616}}.

\bibitem{Haensch:2023sig}
M.~Haensch, F.~Rennecke and L.~von Smekal, \emph{{Medium Induced Mixing and
  Critical Modes in QCD}},  \href{https://arxiv.org/abs/2308.16244}{{\ttfamily
  2308.16244}}.

\bibitem{Winstel:2024dqu}
M.~Winstel, \emph{{Spatially oscillating correlation functions in
  $\left(2+1\right)$-dimensional four-fermion models: The mixing of scalar and
  vector modes at finite density}},
  \href{https://arxiv.org/abs/2403.07430}{{\ttfamily 2403.07430}}.

\end{thebibliography}\endgroup
\end{document}